%% file: main.tex
\documentclass[conference,10pt]{IEEEtran_ICRATmod}

\IEEEoverridecommandlockouts
\usepackage[numbers]{natbib}

\usepackage{graphics} 
\usepackage{epsfig} 
\usepackage{times} 
\usepackage{amsmath} 
\usepackage{amssymb}  
\usepackage{amsthm}
\usepackage{amsfonts}
\usepackage{mathtools}
\usepackage{multirow} 
\usepackage{rotating} 
\usepackage{graphicx}
\usepackage{float}
\usepackage{flushend}
\usepackage[table,xcdraw]{xcolor}
\usepackage{csquotes}
\usepackage{bm}
\usepackage{relsize}
\usepackage{epspdfconversion}
\usepackage{colortbl}
\usepackage{lipsum}
\usepackage[mathscr]{eucal}
\usepackage{fancyhdr}
\usepackage{comment}

\usepackage{tikz}

\usepackage{algorithm}
\usepackage[algo2e]{algorithm2e}

\usepackage{bbm}

\usepackage{hyperref}

\DeclareMathAlphabet\mathbfcal{OMS}{cmsy}{b}{n}

\def\compactify{\itemsep=0pt \topsep=0pt \partopsep=0pt \parsep=0pt}
\newcommand{\compress}{\itemsep=0pt \topsep=0pt \partopsep=0pt \parsep=0pt \leftmargin=30pt \labelwidth=30pt}

\let\latexusecounter=\usecounter

\begin{document}
\title{User Feedback-Informed Interface Design\\for Flow Management Data and Services (FMDS)}

\author{\scriptsize
	   	   \IEEEauthorblockN{Sinan Abdulhak, Anthony Carvette, Kate Shen, Max Z. Li}
	   	   \IEEEauthorblockA{University of Michigan\\Ann Arbor, MI, USA\\ \{\href{mailto:smabdul@umich.edu}{smabdul}, \href{mailto:atcarv@umich.edu}{atcarv}, \href{mailto:kateshen@umich.edu}{kateshen}, \href{mailto:maxzli@umich.edu}{maxzli}\}@umich.edu}
        \and
        \IEEEauthorblockN{Robert Goldman, Bill Tuck}
	   	   \IEEEauthorblockA{Delta Air Lines\\Atlanta, GA, USA\\ \{\href{mailto:Robert.S.Goldman@delta.com}{Robert.S.Goldman}, \href{mailto:Bill.Tuck@delta.com}{Bill.Tuck}\}@delta.com}
}


\vspace{-2cm}
\maketitle

\renewcommand{\headrulewidth}{0pt}
\lhead{\textcolor{black}{ICRAT 2024}}
\rhead{\textcolor{black}{Nanyang Technological University, Singapore}}
\cfoot{\textcolor{black}{\thepage}}
\lfoot{}
\thispagestyle{fancy}
\pagestyle{fancy}

\fancypagestyle{firststyle}
{
    \fancyhf{}
    \lhead{\textcolor{black}{ICRAT 2024}}
\rhead{\textcolor{black}{Nanyang Technological University, Singapore}}
    \cfoot{\textcolor{black}{\thepage}}
    \lfoot{\textcolor{white}{\thepage}\\ \vspace{0.1cm} \scriptsize \parbox{0.465\textwidth}{\hrule ~~\\ S. Abdulhak, A. Carvette, K. Shen, and M. Z. Li acknowledge resources and feedback provided by the 2023 FAA TFM-AID Challenge \cite{TFM_AID_Challenge}. However, this article solely reflects the opinions and conclusions of its authors and not that of the FAA, NASA, Delta Air Lines, or any other private or public entity.}}
}

\noindent
\begin{abstract}
The transition to a microservices-based Flow Management Data and Services (FMDS) architecture from the existing Traffic Flow Management System (TFMS) is a critical enabler of the vision for an Information-Centric National Airspace System (NAS). The need to design a user-centric interface for FMDS is a key technical gap, as this interface connects NAS data and services to the traffic management specialists within all stakeholder groups (e.g., FAA, airlines). We provide a research-driven approach towards designing such a graphical user interface (GUI) for FMDS. Major goals include unifying the more than 50 disparate traffic management services currently hosted on TFMS, as well as streamlining the process of evaluating, modeling, and monitoring Traffic Management Initiatives (TMIs). Motivated by this, we iteratively designed a GUI leveraging human factors engineering and user experience design principles, as well as user interviews. Through user testing and interviews, we identify workflow benefits of our GUI (e.g., reduction in task completion time), along with next steps for developing a live prototype.
\end{abstract}

\begin{small}{{\bfseries\itshape Keywords---Decision support systems; Human-machine collaboration; Human factors; User experience; Traffic Flow Management System (TFMS); Flow Management Data and Services (FMDS)}}\end{small}

\input{ICRAT2024_FMDS/chapters/intro.tex}
\input{ICRAT2024_FMDS/chapters/problem_statement}
\input{ICRAT2024_FMDS/chapters/lit_review}
\input{ICRAT2024_FMDS/chapters/contribution}

\input{ICRAT2024_FMDS/chapters/design_methods}
\input{ICRAT2024_FMDS/chapters/design_decisions}

\input{ICRAT2024_FMDS/chapters/conclusion}

\input{ICRAT2024_FMDS/chapters/acknowledgements}

~~

\bibliographystyle{IEEEtran} 
\small{
\bibliography{main.bib}
}


\end{document}

%% file: ICRAT2024_FMDS/chapters/intro.tex
\thispagestyle{firststyle}
\section{Introduction} \label{sec:intro}

The January 11\textsuperscript{th} Notice to Air Missions (NOTAM) outage put a spotlight on the increasingly aging air traffic management (ATM) and services infrastructure as flight operations across the US were halted through a multi-hour, nationwide ground stop \cite{FAA_NOTAM_2023}. To address this infrastructure gap and acknowledging the increasingly data-driven nature of the US National Airspace System (NAS), the Federal Aviation Administration (FAA) has introduced the \emph{Information-Centric NAS} (ICN) concept \cite{FAA_icn}. ICN builds on previous, large-scale investments such as the Next Generation Air Transportation System (NextGen) program \cite{FAA_nextgen}, and one of the central enabling infrastructure components is the transition from the Traffic Flow Management System (TFMS)---a 13 year-old data exchange system for managing and monitoring the flow of air traffic---to \emph{Flow Management Data and Services} (FMDS) \cite{TFMSC_Manual}. 

\begin{figure}[ht] 
    \centering
    \includegraphics[width=1\linewidth]{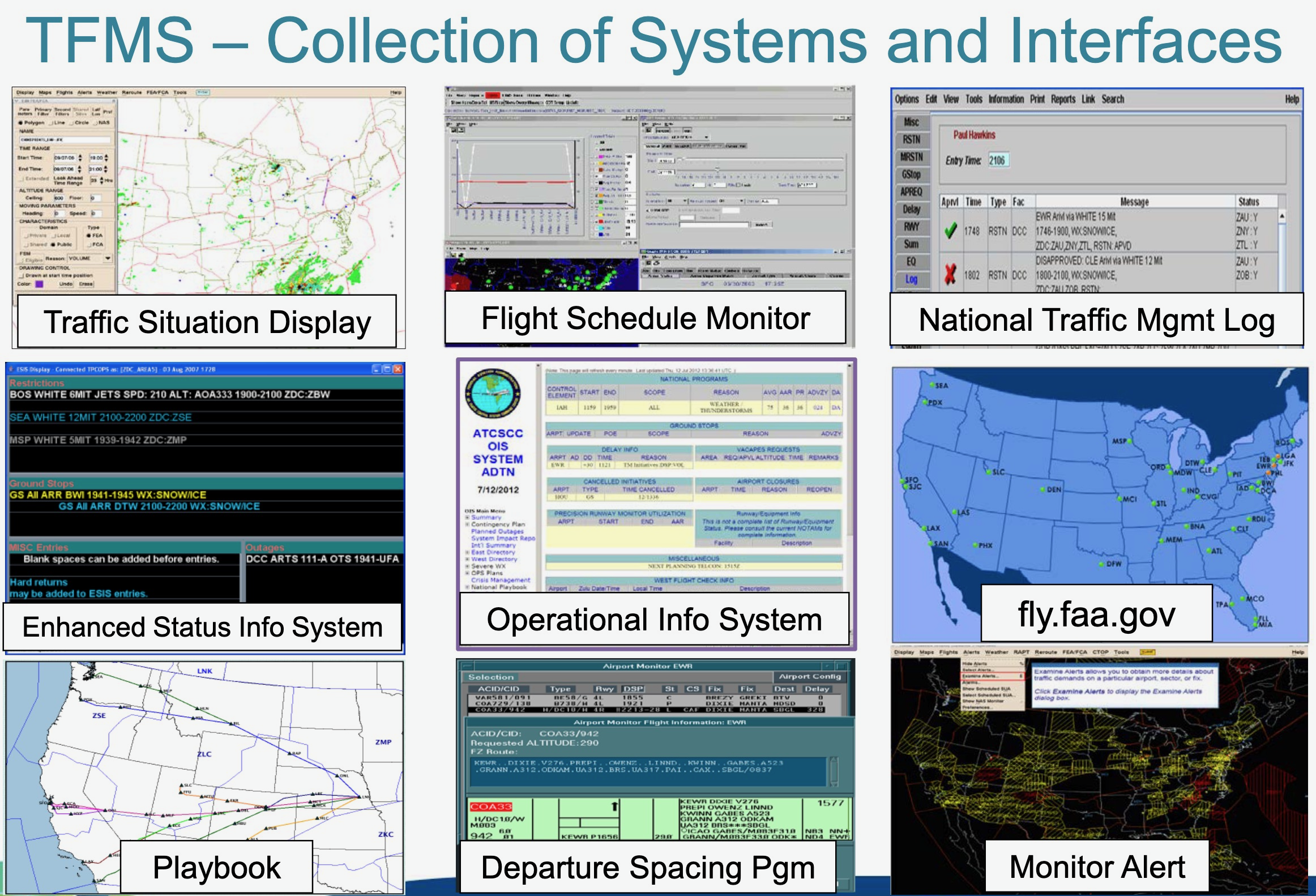}
    \caption{Examples of the more than 50 applications that comprise TFMS \cite{FAA_ATFM}.}
    \label{fig:fig1}
\end{figure}

FMDS will provide increased automation by using a microservices-based software architecture, which translates into decreased response times between NAS stakeholders and increased system reliability and maintainability \cite{FMDS}. With demand for air travel set to double from 2019 to eight billion passengers by 2040, FMDS will enable a corresponding scaling of Traffic Flow Management (TFM) activities to meet rising demand \cite{IATA, FMDS}. Perhaps most importantly for ATM users, FMDS will feature a streamlined, unified interface that air traffic managers will use to perform TFM activities. This is a significant change from the current TFMS experience, where users have to navigate through more than 50 separate applications (Fig. \ref{fig:fig1}) to perform their tasks. This creates a disjointed workflow and results in an inconsistent and inefficient user experience \cite{wickens, FAA_ATFM}.

The TFMS to FMDS infrastructure upgrade will affect a significant portion of the work portfolio within the Air Traffic Organization (ATO) at the FAA, given that TFMS is used in 138 sites with 800 FAA users \cite{FAA_ATFM}. TFMS enables the effective management of air traffic throughout the NAS with Traffic Management Initiatives (TMI) such as Ground Delay Programs (GDPs) and Airspace Flow Programs (AFPs). 
Within the FAA, National Traffic Management Specialists (NTMS) rely on TFMS to coordinate across facilities and approve restrictions, involving Air Route Traffic Control Centers (ARTCC) and Terminal Radar Approach Control Facilities (TRACON) in the decision-making process. TFMS also enables FAA and industry stakeholders (e.g., airlines) to communicate and coordinate on ATM strategies. Through this upgrade, the FAA will need to consider how FMDS---and how NTMS interact with FMDS through critical user interfaces---can improve the significant coordination and collaboration efforts that take place throughout the TFM process \cite{FAA_CDM}.

%% file: ICRAT2024_FMDS/chapters/problem_statement.tex
\subsection{Technical gap and research approach}    \label{sec:bg}



Currently, although TFMS enables traffic management practices and Collaborative Decision-Making (CDM), the TMI creation workflow for NTMS splits user attention and introduces opportunities for error  \cite{wickens, ConOps}. FMDS seeks to unify disparate applications and minimize user training. Based on User Experience Design (UXD) and Human Factors Engineering (HFE) principles, we designed a graphical user interface (GUI) for FMDS, targeting the workflow process of evaluating, modeling, and monitoring TMIs using AFPs as a case example. Our GUI design approach aims to make the NTMS workflow more efficient and facilitate easier collaborations. 

%% file: ICRAT2024_FMDS/chapters/lit_review.tex
\section{Literature review} \label{sec:lit_review}

\subsection{NAS traffic flow management automation}

The role of TFMS and its predecessor, the Enhanced Traffic Management System (ETMS), in coping with the rapid growth of air travel through strategic ATM is well-documented in existing literature. 
To address air travel surges from 1980-1990 and the accompanying workload increases, the FAA developed ETMS in the 1980s, 
with further expansion to ARTCCs and TRACONs past 1992 \cite{goeddel}. The use of ETMS and ETMS-enabled processes (e.g., CDM) yielded many benefits to airlines, passengers, and the FAA, including a reduced number of NAS traffic restrictions, improved controller workloads, and a significant reduction in delay hours (e.g., 49,100 hours saved in 1990) \cite{chang, goeddel}.  
Over the next few years, the FAA developed more tools to assist Traffic Managers in improving their management strategy, such as the Post-Operations Evaluation Tool (POET), and began development of TFMS. TFMS became fully operational in 2010, replacing ETMS and providing benefits such as improved schedule predictability and workload balancing \cite{yoo, reynolds}. 

\subsection{Human factors engineering and user experience design in air traffic automation solutions}

The field of HFE is rooted in decades of research on human psychology and offers guidance on how to design human-machine systems that maximize user safety and satisfaction while minimizing user error. 
To design usable systems, designers must understand user needs, test with users, and iterate on designs \cite{gould, nielsen_a}. Specific to aviation, \cite{wise} concludes that HFE principles should be applied to the design of ATM interfaces to create input devices that map to the displays they control, encourage quick and error free task completion, and provide continuous feedback to the user. UXD further improves interactive interface usability by shifting from viewing users as mere sources of information to recognizing them as key contributors to the overall system experience \cite{Tosi2020}. As part of this shift, UXD experts highlight how emotions evoked by user experiences are key considerations in the UXD process \cite{mccarthy}. 

%% file: ICRAT2024_FMDS/chapters/contribution.tex
\section{Contribution of work}    \label{sec:contribution}
The contributions of our work is as follows:
\begin{itemize}
    \item We document user pain points specific to TFMS-based interfaces through ten interviews with FAA and industry stakeholders, including NTMS, Delta Air Lines (DL), and ATM researchers at the National Aeronautics and Space Administration (NASA);
    \item We design a dual-monitor interface prototype for FMDS that streamlines the TMI workflow through three \textit{Core functionalities} and strengthens information sharing with an internal \textit{Collaboration Feature}, using AFPs as a case example;
    \item We use feedback from FAA and industry stakeholders, HFE theories, and UXD principles to design and evaluate the usability of our GUI.
\end{itemize}

%% file: ICRAT2024_FMDS/chapters/design_methods.tex
\section{Design methods} \label{sec:design_methods}

\subsection{User interviews} \label{sec:User Interviews}
User interviews are a vital, well-established method in HFE and UXD for creating products that satisfy user needs \cite{Lee2017_IHFE}. To this end, we interview the following critical stakeholder groups for FMDS: 

\subsubsection{FAA} \label{sec:FAA}

We interviewed Traffic Managers at the FAA Air Traffic Control System Command Center (ATCSCC, or Command Center), given that these personnel will be the primary users interacting with FMDS. Our team sought the perspectives of Traffic Managers in regions where AFP use is frequent, such as Florida and New York. Interviewees included a Traffic Management Coordinator (TMC) at the FAA Jacksonville ARTCC (ZJX) who formerly worked in the ATCSCC and an TMC with 24 years of experience in New York TRACON (N90), which involves ATCSCC-like decision-making \cite{interview_doc}. 

\subsubsection{Airline ATM teams} \label{sec:Airline}
We interviewed two DL Operations and Customer Center (OCC) personnel, including a current General Manager (GM) of the DL ATM team and a Senior Systems Operations Manager. The interviewed GM has 26 years of experience in DL's OCC, including working as Supervisor of the Strategic Planning Team, which engages with the FAA on a daily basis to coordinate ATM strategy. The interviewed Senior Systems Operations Manager had 14 years of experience at Northwest Airlines and 14 years of experience at DL's OCC, including as a Digital Technology Manager for DL's operations tools \cite{interview_doc}. 

\subsubsection{NASA} \label{sec:NASA}

Our team conducted interviews with three researchers from NASA's ATM group, all of whom have extensive experience in NAS automation technology development and transfer. 
The diverse expertise within this research group, including a Human Factors Engineer, a Data Analytics professional, and a former FAA NTMS provided valuable real-world perspectives on how interfaces and data-centric applications can enhance the efficient execution of ATM \cite{interview_doc}.


\subsection{Design principles} \label{sec:Design Principles}
The overriding principle of HFE and UXD methodologies is that the product (e.g., our GUI), should adapt to users' (e.g., NTMS) needs, rather than forcing users to adapt to the product \cite{pea}. 
Our team relied on these methodologies to develop a unified GUI for NTMS that aims to reduce distractions and enhance their existing workflows.

\subsubsection{Principles of display design} \label{sec:Principles of Display Design}
The \textit{\enquote{Salience Compatibility}} principle states that important and urgent information should attract attention \cite{Lee2017_IHFE}. Information can be made more salient through the use of, e.g., color. The \textit{\enquote{Minimize Information Access Cost}} asserts that there is a cost in time and effort to move a user's selective attention from one display location to another \cite{wickens}. In the existing Traffic Situation Display (TSD), users must navigate through five layers of nested menus to modify a Flow Evaluation Area (FEA) \cite{TSD_CDM}, thus increasing its information access cost. The \textit{\enquote{Proximity Compatibility}} principle, used frequently in the design of avionics, states that information should be arranged such that their mental and display proximities align \cite{wise}. 


\subsubsection{Usability heuristics for user interface design} \label{sec:Usability Heuristics for User Interface Design}

Human factors and usability professionals leverage design heuristics, based on human behavior research, to create systems that account for human capabilities \cite{heuristics,Lee2017_IHFE}. 
With the TFMS to FMDS transition, we had to balance between users accustomed to TFMS, new users with limited TFMS experience, and evolving design trends. Referencing \cite{heuristics}, we aimed to enhance workplace efficiency while ensuring a smooth transition for existing TFMS users, avoiding overwhelming changes in appearance and functionality. Finally, given the highly specialized role of an NTMS with many acronyms and terminology, we relied on the \textit{\enquote{Match between System and the Real World}} principle to ensure the interface adapts to user needs by using familiar words, phrases, and concepts \cite{heuristics}. Furthermore, NTMS often have decades of ATM experience, developing preferences and customizations: The \textit{\enquote{Flexibility and Efficiency of Use}} principle states that customization (e.g., shortcuts, personalized content) should be provided for expert users to help accomplish their tasks more efficiently \cite{heuristics}---for example, several customizable features exist within the Flight Schedule Monitor (FSM) \cite{FSM}.


\subsubsection{Data visualization technical communication principles} \label{sec:Principles of Technical Communication for Data Visualization}
An FMDS GUI should support ATM personnel through intuitive visuals that communicate complex and important data. Within FSM, gridlines and small font sizes can hinder interpretation of demand-capacity graphs. Clutter can be decreased by optimizing the \textit{\enquote{Data-Ink Ratio}} \cite{tufte}---with few exceptions, the most amount of data should be presented with the least amount of content. 
Although we do not explicitly compute such ratios, the general principle guided our design decisions.

%% file: ICRAT2024_FMDS/chapters/design_decisions.tex
\section{Iterative design process} \label{sec:Iterative Design Process}

We regularly revisited and revised designs to incorporate insights gleaned from user interviews. 
In the initial stages of our design process, our primary focus was understanding user pain points. We utilized published FAA resources (see Section \ref{sec:FAA}), explored TFM training materials, and engaged in conversations with current and former TFMS users. 



\begin{figure*}[!htbp] 
    \centering
    \includegraphics[width=1\linewidth]{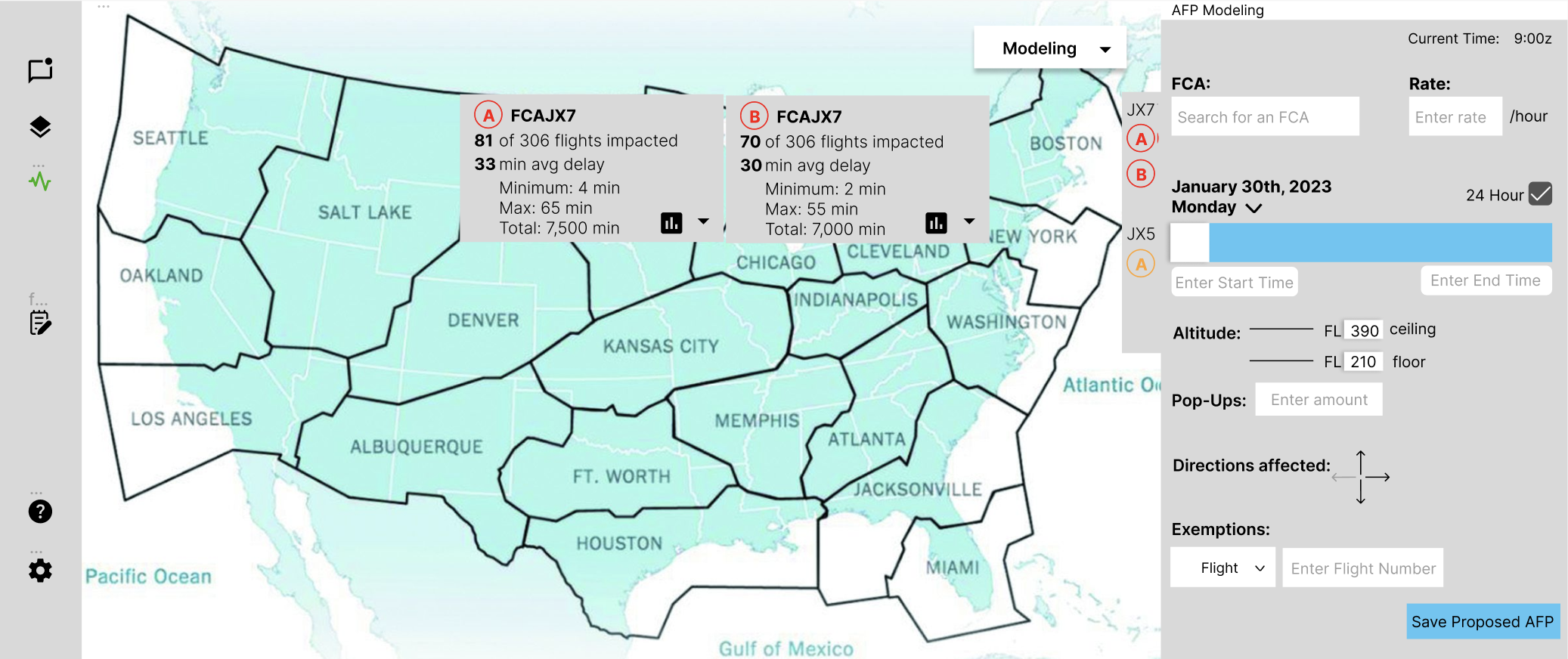}
    \caption{Graphical depiction of the Modeling functionality, including a TSD-style map.}
    \label{fig:fig3}
\end{figure*}

\subsection{Initial design sketches} \label{sec:Initial Design Sketches}
Our team documented the approach TMCs use to address demand-capacity imbalances with AFPs, noting that TMCs must first evaluate the need for an AFP given potential NAS constraints (e.g., weather, rocket launches). Then, TMCs model AFP proposals to manage demand-capacity imbalances by adjusting AFP parameters; modeled proposals are shared via the CDM process and iterated upon as needed. Through this, we distilled five Core functionalities for our design:



\begin{itemize}
    \item \textit{Evaluating}: Evaluate AFP need via weather overlays on an interactive map, visualize demand-capacity imbalances, and recap NAS constraints via a \textit{Constraint Summary};
    \item \textit{Modeling}: Input parameters to model different AFPs, view delay metrics and AFP impact at a glance, quickly determine if a proposed AFP addresses a demand-capacity imbalance, and share AFP models with facilities;
    \item \textit{Creating}: After CDM, schedule or implement the proposed AFP, with an automatic update pushed to National Traffic Management Logs (NTML);
    \item \textit{Monitoring}: Monitor AFPs and other FEAs/FCAs (where FCA stands for Flow Constrained Area) with user-centered viewing modes, track AFP revision history, and revise AFPs as needed;
    \item \textit{Canceling}: Purge an AFP when it is no longer needed, with an update pushed to NTML.
\end{itemize}


\begin{figure}[ht] 
    \centering
    \includegraphics[width=1\linewidth]
    {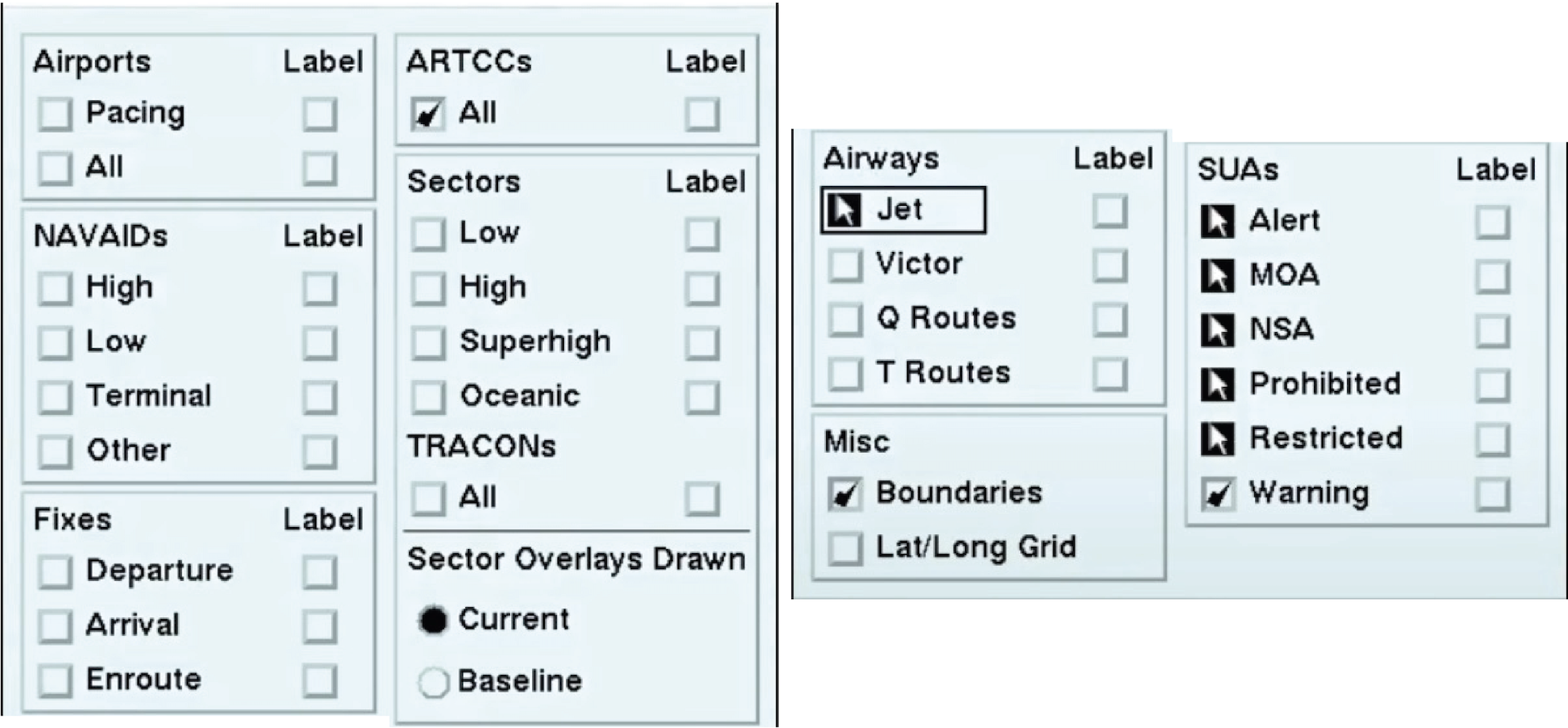}
    \caption{Map overlays available in TFMS \cite{interview_doc}.}
    \label{fig:fig2}
\end{figure}

During an interview with a TMC, we observed their consistent use of the TSD-style map for monitoring traffic flows, and discovered the map's customization capabilities, allowing users to determine visible constraints (Fig. \ref{fig:fig2}) \cite{interview_doc}. Recognizing the importance of map overlays when evaluating the need for an AFP, we incorporated \textit{Layers} into our interface (left of Fig. \ref{fig:fig6}). We also noted that the navigation bar in TFMS had numerous rarely used options \cite{interview_doc}. Thus, we streamlined and prioritized essential functions for TMCs, having our initial \textit{Navigation Bar}, featuring simple icons, placed on the left side of the screen (Fig. \ref{fig:fig3}). This design enhances the user experience by prioritizing relevant and frequently accessed functionalities (see Section \ref{sec:Principles of Display Design}), fostering a more efficient and user-friendly interface for TMCs.

\subsection{Iteration 1: Initial wireframes} \label{sec:Iteration 1}

To start, we translated our most intuitive designs into initial wireframes (Fig. \ref{fig:fig3}) using the prototyping software Figma. 
Figma's built-in tools allowed us to incorporate images of the TSD-style map, add buttons, and implement further iterations and customizations (Fig. \ref{fig:fig3}). 
Additional interviews highlighted the need for TMCs to access more information beyond map visualizations (e.g., the ability to hover over map elements to reveal additional information on traffic flow constraints) \cite{interview_doc}. 
We incorporated the TSD-style map and the concept of a layers feature into our first iteration, empowering TMCs to gain insights into active NAS constants (e.g., rocket launches). 

\begin{figure}[ht] 
    \centering
    \includegraphics[width=1\linewidth]{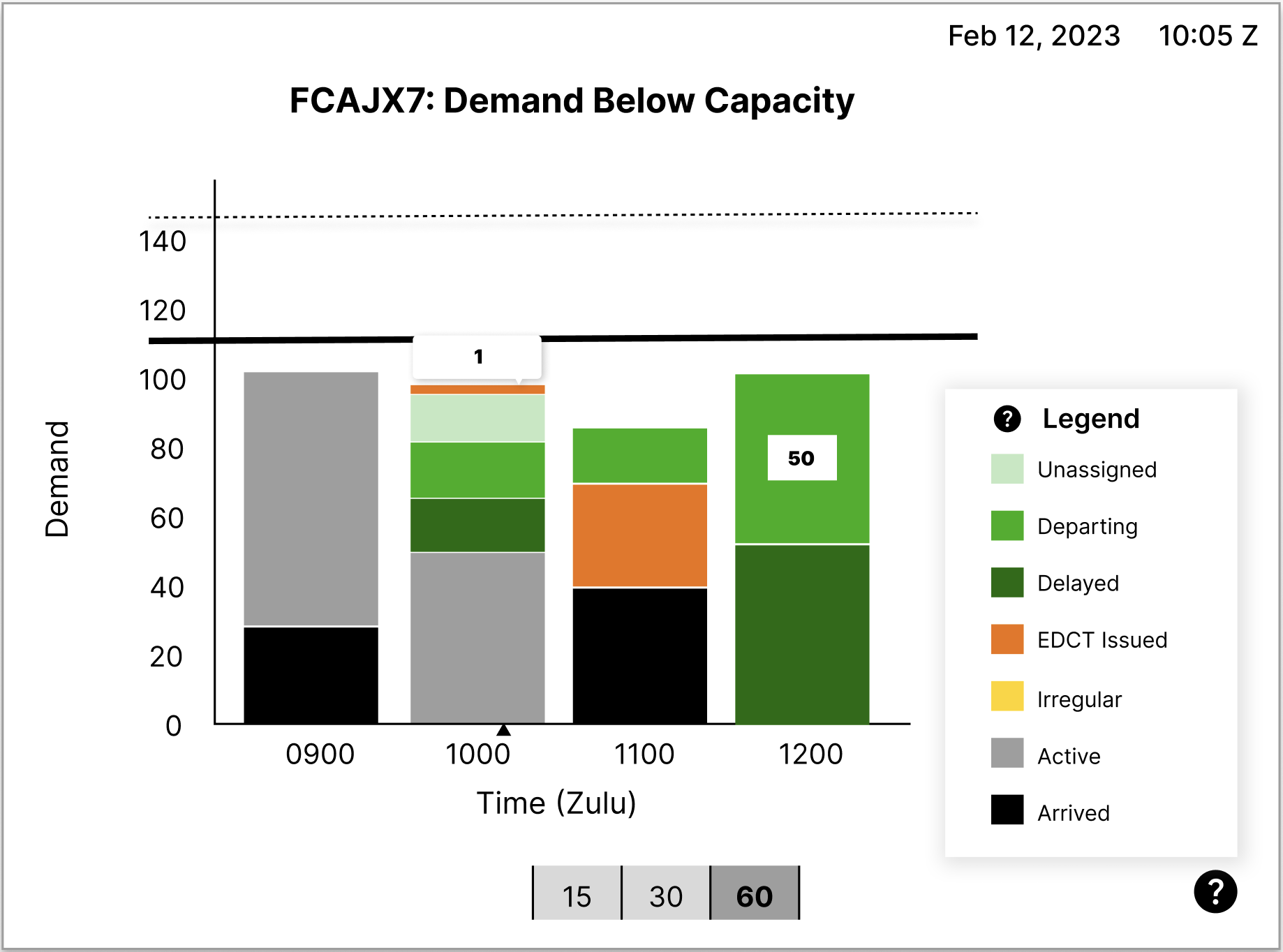}
    \caption{Initial designs for FSM graphs.}
    \label{fig:fig4}
\end{figure}

\begin{figure}[ht] 
    \centering
    \includegraphics[width=1\linewidth]{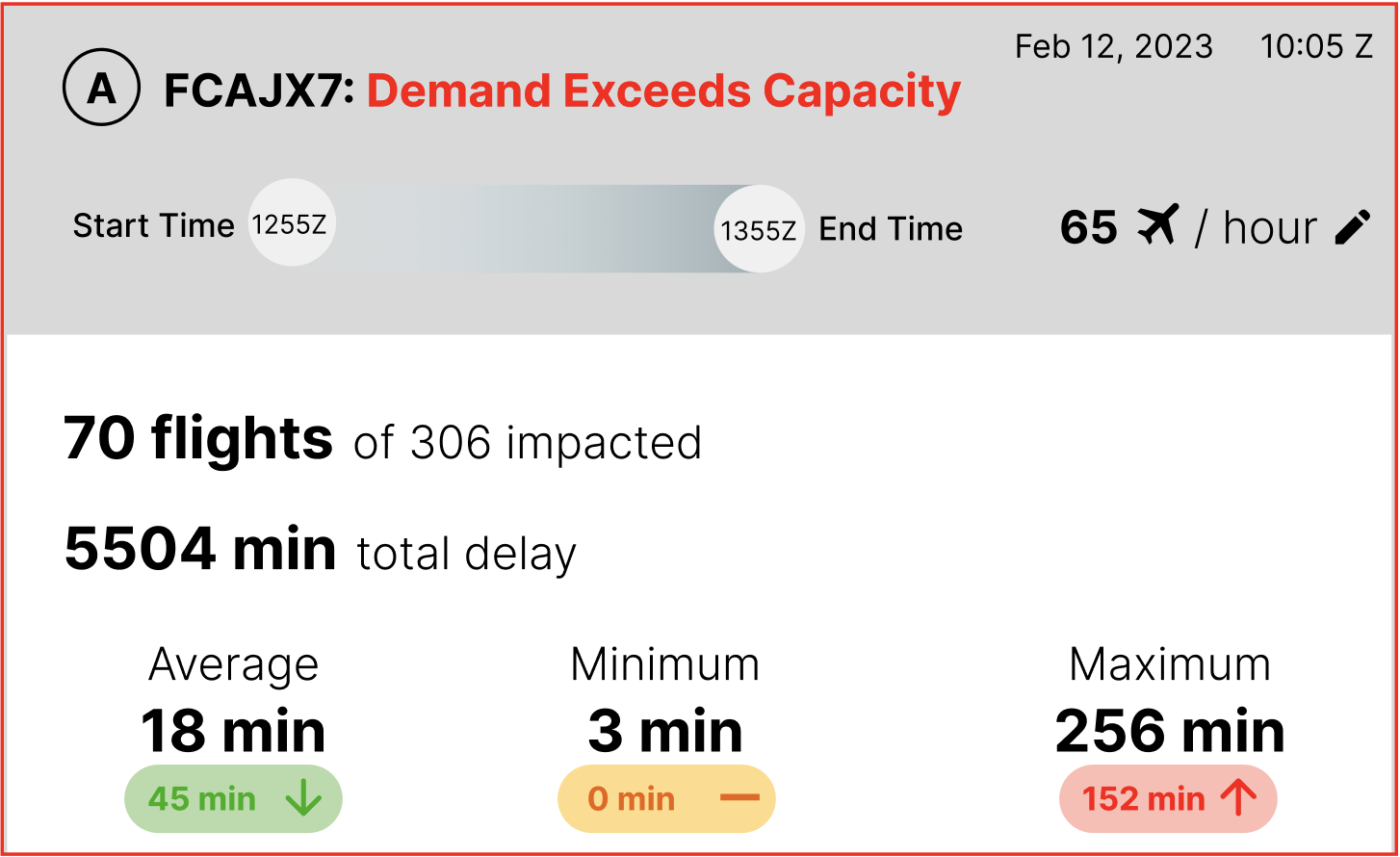}
    \caption{Updated designs for Data Cards.}
    \label{fig:fig5}
\end{figure}

Subsequent interviews with the NASA ATM group revealed the multitude of metrics, such as Expected Departure Clearance Times (EDCT), average delay time, and the number of flights impacted, that TMCs rely on to comprehensively assess AFP proposals \cite{interview_doc}. Recognizing the importance of summarizing these metrics, we introduced the \textit{Data Card} to our GUI, providing NTMS with a consolidated view for efficiently viewing and comparing proposals (Fig. \ref{fig:fig5}). We also integrated FSM graphs (Fig. \ref{fig:fig4}) into our Data Cards, as TMCs emphasized their usefulness in identifying demand-capacity imbalances \cite{interview_doc}. In line with optimizing the \enquote{Data-Ink Ratio} and \enquote{Salience Compatibility} principles (see Sections \ref{sec:Principles of Technical Communication for Data Visualization} and \ref{sec:Principles of Display Design}), we removed gridlines, increased font sizes, and used colors to correspond with various levels of urgency (e.g., gray for flights that are ineligible for delay assignment).


\subsection{Iteration 2: Improving user experience through familiarity} \label{sec:Iteration 2}

\begin{figure}[ht] 
    \centering
    \includegraphics[width=1\linewidth]{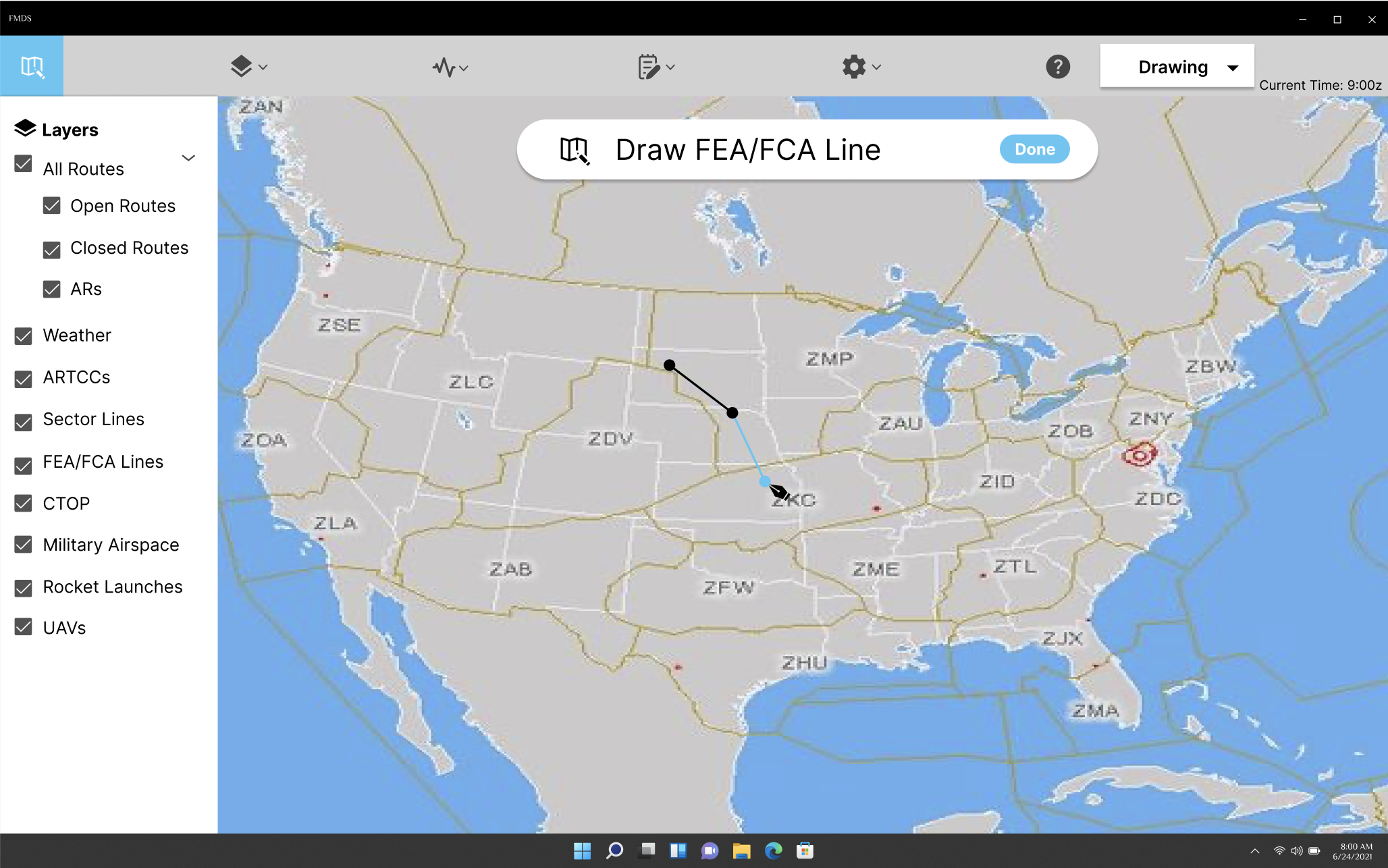}
    \caption{Graphical depiction of  the Drawing Tool for the Evaluating functionality.}
    \label{fig:fig6}
\end{figure}

A significant change we implemented was relocating our Navigation Bar to the top of the interface (Fig. \ref{fig:fig6}) for improved functionality. 
Recognizing the diverse range of NTMS experience (from novices to those with decades of TFMS usage), we prioritized functionality and efficiency in our GUI, making it familiar to all users by emulating other interfaces they may be acquainted with (see Section \ref{sec:Usability Heuristics for User Interface Design}). Our Collaboration Feature functions similarly to Microsoft Teams \cite{microsoft_teams}, with chat threads organized by group or topic, and the option to conduct voice calls. Adhering to the \enquote{Match between System and the Real World} principle (see Section \ref{sec:Usability Heuristics for User Interface Design}), the Navigation Bar is placed on top of our GUI, identical to its placement within TFMS-based applications such as TSD \cite{TSD}. To further enhance familiarity for NTMS, a \textit{Drawing Tool} was developed for the Evaluating functionality (Fig. \ref{fig:fig6}), incorporating feedback from the NASA ATM group and real-life demonstrations of how TMCs assess demand by drawing FEA lines \cite{interview_doc}.

Due to multiple daily conferences (e.g., Advanced Planning Webinar) and the collaborative nature of addressing demand-capacity imbalances,  
we recognized the need for a local collaboration feature within FMDS, allowing for chat and voice communication \cite{interview_doc}. We aim to enhance the CDM process through integrating these virtual meetings into our GUI, as NTMS no longer need to use myriad communication platforms (e.g., phone calls, email, Microsoft Teams). The use of a centralized communication platform aligns with the \enquote{Minimize Information Access Cost} principle (see Section \ref{sec:Principles of Display Design}).



\begin{figure*}[!htbp] 
    \centering
    \includegraphics[width=1\linewidth]{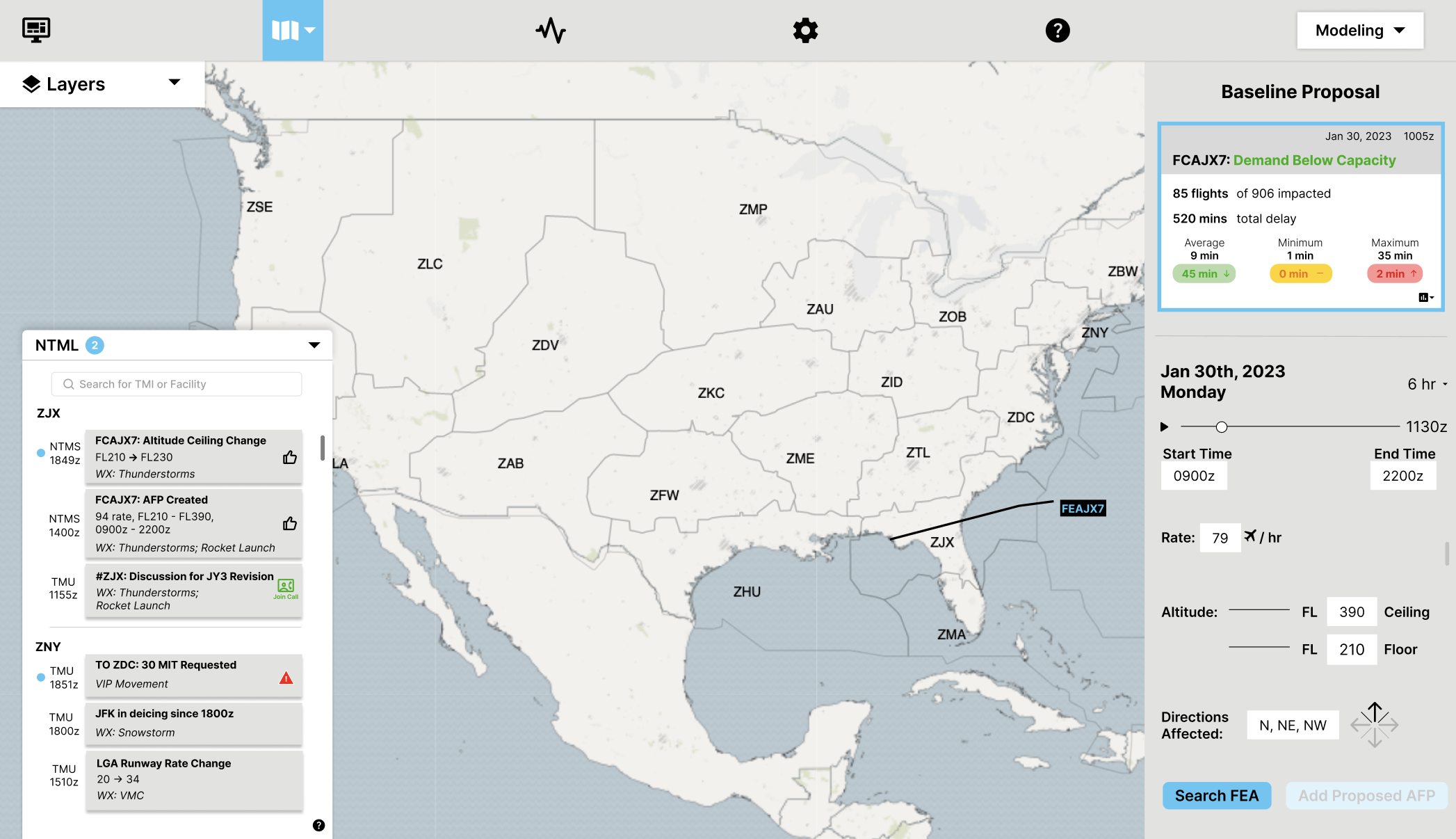}
    \caption{Final iteration of the GUI as it appears on the left monitor, depicting the Monitoring functionality.}
    \label{fig:fig7}
\end{figure*}

\begin{figure*}[!htbp] 
    \centering
    \includegraphics[width=1\linewidth]{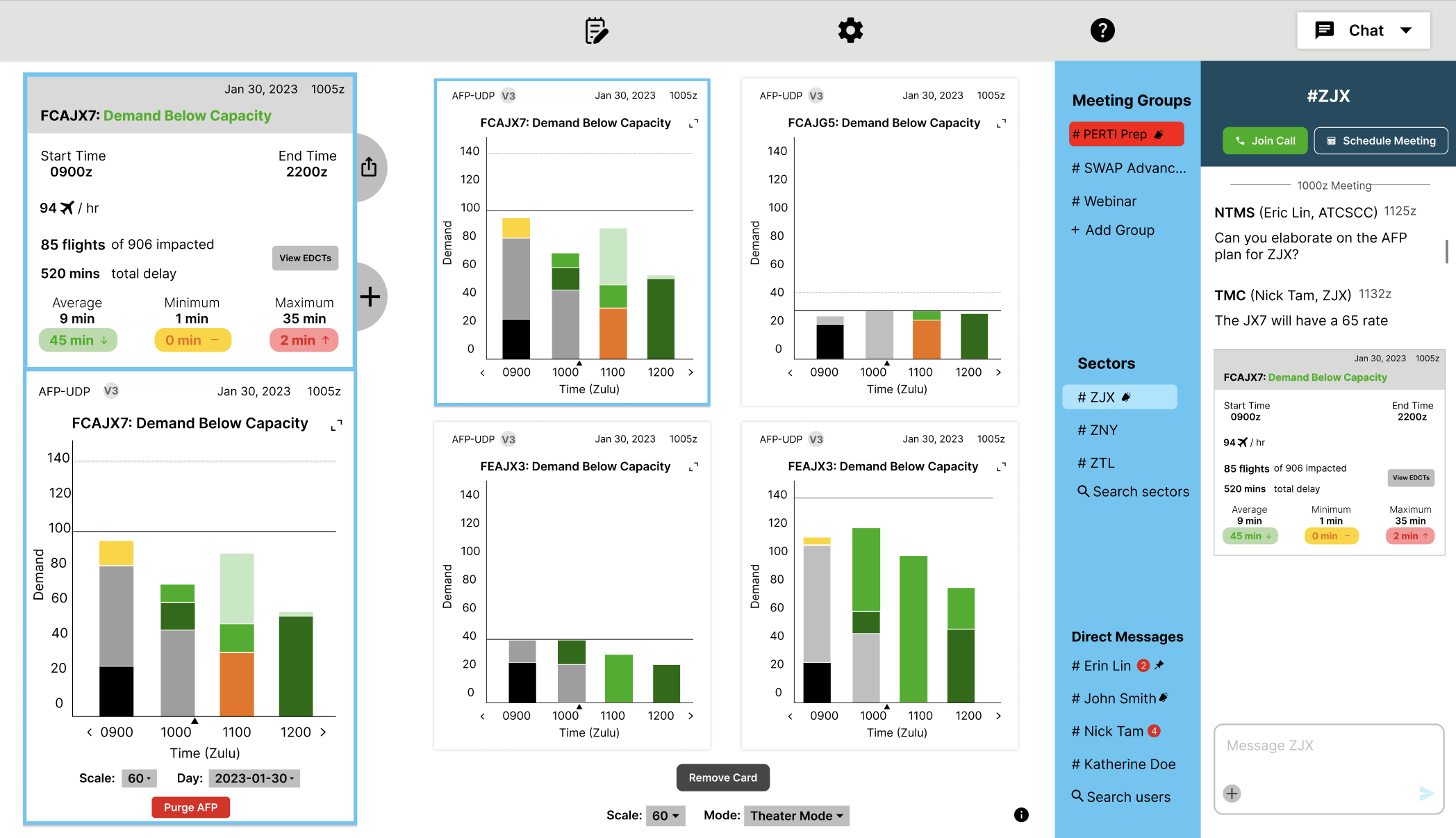}
    \caption{Final iteration of the GUI as it appears on the right monitor, complementary to Fig. \ref{fig:fig7}.}
    \label{fig:fig8}
\end{figure*}

To better streamline the numerous services hosted within FMDS, we adopted a dual-monitor setup (Fig. \ref{fig:fig7} and Fig. \ref{fig:fig8}). This was also driven by interviews with the NASA ATM group, where we noted that all ATCSCC positions use at least two screens, with NTMS commonly using three to five monitors \cite{interview_doc}. The left monitor (Fig. \ref{fig:fig7}) focuses on identifying demand-capacity imbalances, where NTMS interact with the TSD-style map and create AFP proposals. The right monitor (Fig. \ref{fig:fig8}) is dedicated to CDM-related functionalities, allowing NTMS to compare several Data Cards simultaneously. 


\subsection{Iteration 3: Core functionalities and CDM} \label{sec:Iteration 3}
Until this iteration, our interface was comprised of five Core functionalities, designed to guide NTMS through the entire AFP workflow from evaluating AFP need to purging an AFP (see Section \ref{sec:Initial Design Sketches}). In this iteration, we consolidated down to three functionalities, as we found similarities among the Core functionalities. After consolidation, we retained  \textit{(i) Evaluating} AFP need; \textit{(ii) Modeling} an AFP proposal; and \textit{ (iii) Monitoring} an active AFP. Users now only have to remember and choose between three Core functionalities, decreasing cognitive load.

To enhance CDM, our Collaboration Feature and redesigned \textit{NTML} aim to centralize communication and capture users' attention with effective notifications. 
We incorporated a pulsing bar in the chat thread header for important messages (Fig.  \ref{fig:fig8}), in line with previous research on air traffic notifications \cite{imbert}, as well as the \enquote{Salience Compatibility} principle (see Section \ref{sec:Principles of Display Design}). To prevent notification fatigue, we default to \textit{mute} chat notifications, allowing users to \textit{emphasize} important threads with the pulsating bar. 

During user interviews, we noted that NTMS communications are spread across multiple platforms (see Section \ref{sec:Iteration 2}); some prefer to converse over phone, while other conversations are better suited for chat (see Section \ref{sec:STMC}). Our Collaboration Feature centralizes NTMS communications, thus removing the need to navigate to multiple platforms. 
Each thread contains chat messaging, where users can share Data Cards (Fig. \ref{fig:fig8}), and includes a voice room for discussions, which is always active. Users can schedule structured meetings via the \textit{Schedule Meeting} button or a preset \texttt{/schedule-meeting} hotkey, enabling users to select their optimal preference. This customization option aligns with the \enquote{Flexibility and Efficiency of Use} principle (see Section \ref{sec:Usability Heuristics for User Interface Design}). 

\subsection{User tests} \label{sec: User Tests}

Once we built an interactive prototype, our team conducted user tests, gathering feedback from experienced TFMS users interacting with our GUI. By assigning tasks and observing users' interactions, we identified the strengths and weaknesses of our GUI and gauged interpretability of the interface through recorded time trials. Despite users seeing our interface for the first time and being given minimal instructions, users completed each task (e.g., create an AFP in ZJX) in an average of 8.07 seconds---for context, previous work has found that working memory decays significantly after ten seconds \cite{interview_doc, wise}. Our interface allows users to accomplish certain tasks in less clicks when compared to TFMS. For example, a TFMS subject matter expert stated that creating an FCA and assigning it a rate are done in entirely separate TFMS programs, where the latter task is far more time consuming \cite{interview_doc}. In our GUI, users can enter the rate in the same place where the FCA is created, translating into less button clicks and a simpler workflow. 

\subsubsection{Traffic Management Coordinator perspectives} \label{sec:STMC}

Our initial user test featured a former TMC at N90, one of the busiest TRACONs in the country \cite{AFP_use}. The former TMC, who has been a TFMS user since it was first developed, completed each task using our GUI in an average of 9.44 seconds. 
They found that our use of recognizable icons facilitated task completion with minimal guidance, aligning with the \enquote{Match between System and the Real World} principle (see Section \ref{sec:Usability Heuristics for User Interface Design}). 
The user test, comprising of 14 tasks, highlighted the utility of features like 
automated NTML logging, where the former TMC emphasized how their experience using our interface is better than TFMS. They mentioned potential hesitancy towards relying on a chat feature, favoring verbal communication instead \cite{dennis}. However, they indicated this sentiment could change with more user feedback, as some conversations are better suited for chat (e.g., discussing a decision, where documentation is desirable). We provide notes from user tests in \cite{interview_doc}.

\subsubsection{Airline perspectives} \label{sec:Airlines}

To evaluate the usability of our GUI for airlines, we conducted a user test with two DL OCC employees, each of whom have nearly 20 years of experience with TSD and FSM. The two employees successfully completed 38 tasks with minimal assistance. The former OCC Digital Technology Manager expressed that the dual-monitor interface was intuitive, facilitating task completion. Both participants accomplished tasks in an average time of 6.28 seconds, attributing their ability to successfully complete tasks to the use of familiar icons. The GM of ATM at DL highlighted how our interface's one-click \textit{Theater Mode} (Fig. \ref{fig:fig8}), where one FSM graph is enlarged, resembled how they currently view and manually arrange FSM boards. Additionally, the former OCC Digital Technology Manager drew parallels between our \textit{Dashboard}, a feature where TMCs view NAS-wide initiatives, aviation weather updates, and upcoming meetings with a calendar tool, and other FAA applications (e.g., how the Flight Planning System works with a TSD-style map). 

\subsection{Final iteration} \label{sec:Iteration 4}

In this iteration, we first focused on display improvements in line with design principles such as \enquote{Minimize Information Access Cost} (by improving figure readability and contrast) and \enquote{Flexibility and Efficiency of Use} (by adding an ability to remove Data Cards and customize what is displayed on screen). We also sought to combine these principles (e.g., making NTML easier to access and control its presence on-screen) to facilitate the user experience.




From our interviews, users expressed strong appreciation for our Data Cards, as it facilitates interpretation of important metrics. We found that users had to split their attention across the two monitors, as information from the Data Cards (on the right monitor) is often used in conjunction with the TSD-style map (on the left monitor). To resolve this, we referenced the \enquote{Knowledge in the World} principle and the \enquote{Proximity Compatibility} principle,  displaying the selected Data Card adjacent to the TSD-style map (Fig. \ref{fig:fig7} and Fig. \ref{fig:fig8}). This enables users to interpret which FEA/FCA the delay metrics are associated with, creating display proximity by placing two sources of information that need to be mentally integrated close to each other (see Section \ref{sec:Principles of Display Design}). Using this finalized iteration of our GUI, a video demonstration depicting AFP creation within the context of FMDS can be viewed in \cite{video}.



%% file: ICRAT2024_FMDS/chapters/conclusion.tex
\section{Conclusion and future work} \label{sec:conclusion}

Creating an interface for FMDS rooted in user feedback offers the potential to significantly improve user performance and satisfaction while positively influencing ATM strategy. We aimed to design a user feedback-informed GUI by unifying more than 50 disparate TFMS applications to streamline ATM workflows and reduce user onboarding times. In addition to leveraging interviews with FAA and industry stakeholders, we adhered to UXD and HFE principles as part of an iterative design approach. This iterative approach allowed us to continually gather and analyze user feedback from interviews and tests. Future work of immediate value would be a live prototype with real-time NAS data. Additional study limitations include potential geographic biases due to the specific subject matter experts we interviewed; future work should broaden the scope of user experiences considered, particularly when working towards real-world FMDS deployment. Finally, additional future research directions include expanding our GUI to handle other TMI types, incorporating accessibility standards such as the Web Content Accessibility Guidelines (WCAG) test \cite{WCAG}, and addressing edge cases (e.g., NTMS only having access to one monitor) not discovered or discussed during the initial iterations.

%% file: ICRAT2024_FMDS/chapters/acknowledgements.tex
\section*{Acknowledgements} \label{sec:ack}

The authors thank Dr. Yili Liu for his helpful comments and feedback which greatly improved this manuscript.